\documentclass[12pt]{article}

\usepackage{amsfonts}
\usepackage{epsfig}
\usepackage{latexsym}
\usepackage{graphicx}
\def\bequ{\begin{equation}}
\def\eequ{\end{equation}}
\def\barr{\begin{array}}
\def\earr{\end{array}}
\def\half{{1\over 2}}
\def\ben{\begin{equation}}
\def\een{\end{equation}}
\def\bena{\begin{eqnarray}}
\def\eena{\end{eqnarray}}

\def\spa#1{\phantom{\fbox{\rule[-#1cm]{0cm}{0cm}}}}

\def\b1{e^0}

\newcommand{\be}{\begin{equation}}
\newcommand{\ee}{\end{equation}}
\def\bea{\begin{eqnarray}}
\def\eea{\end{eqnarray}}

\def\del {\partial}
\def\nn{\nonumber}

\def\half {{1 \over 2}}

\def\be{\begin{equation}}
\def\ee{\end{equation}}
\def\bea{\begin{eqnarray}}
\def\eea{\end{eqnarray}}

\def\lesssim{\mathrel{\hbox{\rlap{\hbox{\lower4pt\hbox{$\sim$}}}\hbox{$<$}}}}
\def\gtrsim{\mathrel{\hbox{\rlap{\hbox{\lower4pt\hbox{$\sim$}}}\hbox{$>$}}}}

\renewcommand{\thefootnote}{\fnsymbol{footnote}}
\begin{document}

\hfuzz=100pt
\title{{\Large \bf{Janus within Janus}}}
\author{\\Shinji Hirano\footnote{hirano@nbi.dk}
  \spa{0.5} \\
{{\it The Niels Bohr Institute}}
\\ {{\it Blegdamsvej 17, DK-2100 Copenhagen}}
\\ {{\it Denmark}}}
\date{March, 2006}

\maketitle
\centerline{}

\begin{abstract}
We found a simple and interesting generalization of the
non-supersymmetric Janus solution in type IIB string theory. 
The Janus solution can be thought of as a thick $AdS_d$-sliced domain
wall in $AdS_{d+1}$ space. 
It turns out that the $AdS_d$-sliced domain wall can support its own
$AdS_{d-1}$-sliced domain wall within it.
Indeed this pattern persists further until it reaches the
$AdS_2$-slice of the domain wall within self-similar 
$AdS_{p\mbox{ }(2<p\le d)}$-sliced domain walls.
In other words the solution represents a sequence of little Janus
nested in the interface of the parent Janus 
according to a remarkably simple ``nesting'' rule. 
Via the AdS/CFT duality, the dual gauge theory description is in
general an interface CFT of higher codimensions.
\end{abstract}

\renewcommand{\thefootnote}{\arabic{footnote}}
\setcounter{footnote}{0}
\section{Introduction}

The non-supersymmetric backgrounds in string theory are often hard to
control and it is not easy to make any definite statements with 
quantitative precision.
Yet we would eventually have to deal with them, 
as our universe is not supersymmetric.
Also for the purpose of understanding the confinement via the 
gauge theory/gravity duality, we hope to ultimately get a good handle
on the non-supersymmetric circumstances.

A non-supersymmetric solution with reasonable tractability was found
in \cite{Bak:2003jk} in the context of the AdS/CFT duality 
\cite{Maldacena:1997re, Gubser:1998bc, Witten:1998qj, Aharony:1999ti}. 
It is a dilatonic deformation of $AdS_5\times S^5$ in type IIB
supergravity and named Janus due to its rather curious characteristic.  
Several dilatonic deformations exist
\cite{Kehagias:1999tr, Gubser:1999pk, Nojiri:1999sb, Bak:2004yf}, 
but they typically lead to singular geometries. 
Although these solutions exhibit interesting phenomena suggesting
possible gravity duals of confinement, the validity of their
analyses is not too clear.
In contrast the Janus solution is nonsingular, 
and the scalar curvature and string coupling 
can be kept small everywhere in spacetime. 
Furthermore the stability of the Janus solution was partially shown
against the scalar field perturbations in \cite{Bak:2003jk}
and remarkably later in \cite{Freedman:2003ax} 
against a broad class of perturbations, suggesting that it is indeed
stable. 

Perhaps what makes Janus possibly interesting is its remarkable
simplicity. 
Indeed in the dual gauge theory side, the Janus simply corresponds to
having the different SYM coupling in each half of the boundary
spacetime \cite{Bak:2003jk}.  
In other words, the SYM coupling jumps discontinuously when it moves
from one half of the space to another, dividing the boundary spacetime 
into two characterized by two different values of the coupling
constants -- hence the name Janus, the god 
of gates, doors, doorways, beginnings, and endings
in Roman mythology who is often symbolized by two faces.
Two faces are joined at the interface. Although the 4-dimensional
conformal symmetry $SO(2,4)$ is partially broken, the conformal
symmetry $SO(2,3)$ on the interface is preserved. 
Hence the dual gauge theory is thought of as an interface 
CFT.\footnote{An interface is similar to a
  defect. However, as emphasized in
  \cite{Clark:2004sb,D'Hoker:2006uv}, the former does not support any
  new degrees of freedom independent of those in the bulk, while the
  latter does. So we adopt the terminology interface CFT rather than
  defect CFT.}  
Remarkably the supergravity predictions were firmly confirmed by the
dual interface CFT computations \cite{Clark:2004sb}.

Another aspect of the Janus solution is in its relation to the $AdS_4$
domain wall in $AdS_5$ \cite{Karch:2000ct} (see also an earlier
work on the $AdS_4$ domain wall in 5D gauged supergravity
\cite{Behrndt:2002ee}).  
By construction, the Janus
can be thought of as a thick $AdS_4$-sliced domain wall in $AdS_5$.
In \cite{Freedman:2003ax} the Janus was generalized and
studied by making use of the fake supergravity from the domain wall
perspective. Some other aspects of the Janus as a domain wall were
discussed in \cite{Sonner:2005sj}.
The brane configuration for the $AdS_4$ domain wall of
\cite{Karch:2000ct} was proposed in \cite{Karch:2001cw}.
In a similar spirit, one might hope to find a brane configuration for
the Janus solution. The non-supersymmetric Janus clearly does not have
any source or charges, so it does not seem to have such an
interpretation. 
This lead to a search for the supersymmetric Janus, 
and in fact the supersymmetric generalization was found 
in \cite{Clark:2005te} at the level of the 5D gauged supergravity and 
more recently in \cite{D'Hoker:2006uu} in the full type IIB supergravity.

In this paper we study a generalization of the non-supersymmetric
Janus in a different flavour.\footnote{Recently yet another
  possibility of the Janus-type solution was discussed in 
  \cite{Bak:2006nh} concerning the holographic description of the
  cosmological backgrounds.}   
The generalization is bound to complicate the original system.
However, here it will be made with a remarkable simplicity for our new 
solution.  
The new solution in general represents a sequence of little
Janus nested within the parent Janus -- $AdS_2$-sliced Janus $\subset$  
$AdS_3$-sliced Janus $\subset$ $AdS_4$-sliced Janus. 
This generalization follows a remarkably simple ``nesting'' rule. 
Indeed it is almost as simple as the original Janus.

In section 2, we review the Janus solution of \cite{Bak:2003jk}
in type IIB supergravity. 
In section 3, we construct our new solution and propose its gauge
theory dual description. 
In section 4, we end with brief discussions.

\section{A review of the Janus solution}
We focus on the simplest case of the Janus solution in type IIB string
theory --  a dilatonic deformation of $AdS_5\times S^5$ found in
\cite{Bak:2003jk}.  
In this case only the metric $g_{MN}$, 5-form field strength $F_5$, and 
dilaton $\phi$ are activated. Then the equations of motion to be
solved are  
\bea
&&R_{MN}=\half\del_M\phi\del_n\phi+{1\over
  96}F_{MPQRS}F_N^{\hspace{0.15cm}PQRS}\ ,\\
&&\del_M\left(\sqrt{g}g^{MN}\del_N\phi\right)=0\ .
\eea
The ansatz for the Janus solution takes the form
\bea
ds^2&=&f(\mu)\left(d\mu^2+ds_{AdS_4}^2\right)+ds_{S^5}^2\ ,\nn\\
\phi&=&\phi(\mu)\ ,\\
F_5&=&4f(\mu)^{5/2}d\mu\wedge d\omega_{AdS_4}+4d\omega_{S^5}\ .\nn
\eea
The five sphere $S^5$ is intact, so is the $SO(6)$ isometry.
Since the $AdS_4$ space has the isometry $SO(2,3)$, the ansatz
thus respects the $SO(2,3)\times SO(6)$ isometry.
When $\phi(\mu)$ is the constant, the scale factor
$f(\mu)$ will be uniquely determined to be $1/\cos^2\mu$ 
and the geometry is the $AdS_4$ slicing of $AdS_5\times S^5$.  
Once $\phi(\mu)$ starts varying, $f(\mu)$ will deviate from
$1/\cos^2\mu$, yielding the deformation of the $AdS$ space. 
So in general the geometry can be viewed as a kind of 
$AdS_4$-sliced domain wall. 

The equation of motion for the dilaton can be readily solved as
\be
\phi(\mu)'={c_0\over f(\mu)^{3/2}}\ ,
\ee
where the dash $'$ denotes the $\mu$-derivative.
Then the Einstein equations yield
\bea
2f'f'-2ff''&=&-4f^3+{c_0^2\over 2f}\ ,\nn\\
f'f'+2ff''+12f^2&=& 16f^3\ .
\eea
These are equivalent to the motion of a particle governed by the
Hamiltonian
\be
H(f,f')\equiv \half f'f' + V(f)
=\half f'f'-\half\left(4f^3-4f^2+{c_0^2\over 6f}\right)\
, 
\label{hamiltonian}
\ee
with zero energy $H(f,f')=0$.

\begin{figure}[ht!]
\centering \epsfysize=2cm
\includegraphics[scale=0.7]{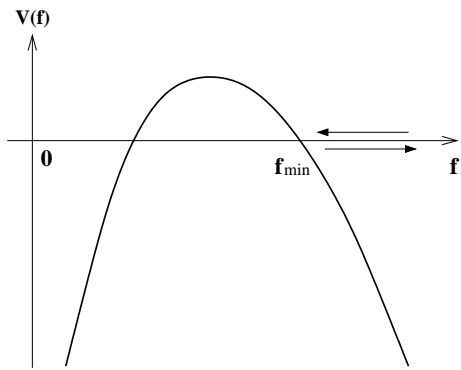}
\caption{The motion of a particle governed by the Hamiltonian
  (\ref{hamiltonian}) with zero energy. 
This motion corresponds to a non-singular asymptotically $AdS$
geometry. As $f$ goes to the infinity and the $c_0$ dependent effect
of the nontrivial dilaton becomes negligible,  
the geometry approaches to the $AdS$ space, 
}
\label{potential}
\end{figure}

There are two possible branches of the solution specified by two
distinct boundary conditions. In Figure \ref{potential}, the potential
of the particle motion is depicted. 
The particle can reach either at $f=0$ or $f=+\infty$.
These two correspond to two different branches of the solution.
However, the former turns out to be a singular solution, since the
scalar curvature diverges at $f=0$.
So we will not be interested in this case.
On the other hand, for the latter $f$ is bounded from below as shown
in Figure \ref{potential}, and the corresponding geometry is
singularity free. 
Furthermore as $f$ goes to the infinity, the $c_0$ dependent effect of
the nontrivial dilaton becomes negiligible. Thus the geometry
asymptotes to the $AdS$ space and
this is the solution we are interested in.

The equation $H(f,f')=0$ can be easily integrated to
\be
\mu=\pm\int^f_{f_{min}}\frac{d\tilde{f}}
{2\sqrt{\tilde{f}^3-\tilde{f}^2+{c_0^2\over 24\tilde{f}}}}\ ,
\label{mu}
\ee
where $f_{min}$ is the largest root of the rational function 
$P(x)=x^3-x^2+{c_0^2\over 24x}$, and we have set the origin of $\mu$
such that $\mu=0$ at $f=f_{min}$ for the convenience.

The dilaton equation can also be integrated to
\be
\phi(\mu)=\phi_0\pm\int^{f(\mu)}_{f_{min}}\frac{c_0d\tilde{f}}
{2\tilde{f}^{3/2}\sqrt{\tilde{f}^3-\tilde{f}^2+{c_0^2\over
      24\tilde{f}}}}\ ,  
\label{dilaton}
\ee
where the choice of the sign is $+$ for $\mu\ge 0$ and $-$ for $\mu<0$.  

The analytic form of the function $f(\mu)$ was recently found in
\cite{D'Hoker:2006uu} in terms of the Weierstrass $\wp$-function.
However here we will focus only on the qualitative features of the
Janus solution. 

First note that as we increase the value of the constant $c_0$, the
potential $V(f)$ goes down. It is easy to show that at
$c_0=9/4\sqrt{2}$ the top of the potential is at $V(f)=0$, 
in other words two roots of the rational function $P(x)$ coalesce. 
Beyond this point, the particle coming in from $f=+\infty$ always
reaches at $f=0$ in the end. So the solution becomes singular for
$c_0>9/4\sqrt{2}$. Hence we will only consider the case 
\be
0\le c_0\le 9/4\sqrt{2}\ .
\ee

Second, since the constant $\mu$ corresponds to the $AdS_4$ slicing of
the (deformed) $AdS_5$, it ranges from $-\pi/2$ to $\pi/2$ in
the undeformed case $c_0=0$. In general $\mu$ is a monotically
increasing function of $c_0$, as one can easily see it from the
expression (\ref{mu}), so its range is
\be
-\mu_0\le\mu\le\mu_0\qquad \left(\mu_0\ge {\pi\over 2}\right)\ .
\ee
Indeed $\mu_0$ diverges when $c_0$ approaches the critical
value $9/4\sqrt{2}$, since it takes an infinite time for a particle to
get to  $f=f_{min}$ when the top of the potential is
precisely level with $V=0$.

Third, the dilaton approaches the constants at $\mu=\pm\mu_0$, as
$f=+\infty$ there and the solution asymptotes to $AdS_5\times S^5$.
In fact the dilaton takes different values, $\phi=\phi_0-\Delta\phi_0$
at $\mu=-\mu_0$ and $\phi=\phi_0+\Delta\phi_0$ at $\mu=\mu_0$, as one
can see it from  
\be
2\Delta\phi_0=
\phi(\mu_0)-\phi(-\mu_0)
=2\int^{\infty}_{f_{min}}{c_0df\over 2f^{3/2}\sqrt{f^3-f^2+{c_0^2\over
      24f}}}>0\ .
\ee

Finally the Janus solution has a peculiar structure at the boundary of
the deformed $AdS_5$. This is where its name comes from. 
To see it, let us first consider the undeformed $AdS_5$ space. 
The $AdS_4$ slicing can be expressed as
\be
ds^2={1\over \cos^2\mu}\left(d\mu^2+ds_{AdS_4}^2\right)\ ,
\ee
where $-\pi/2\le\mu\le\pi/2$.
For the global $AdS_4$ coordinate, it takes the form
\be
ds^2={1\over \cos^2\mu\cos^2\eta}\left(
-d\tau^2+\cos^2\eta d\mu^2+d\eta^2
+\sin^2\eta d\Omega_2^2\right)\ , 
\label{globalAdS4}
\ee
where $0\le\eta\le\pi/2$, 
and for the $AdS_4$ Poincare patch, 
\be
ds^2={1\over y^2\cos^2\mu}\left(
-dt^2+d\vec{x}_2^2+dy^2+y^2d\mu^2\right)\ ,
\label{poincareAdS4}
\ee
where $y\ge 0$.

The former (\ref{globalAdS4}) can be transformed to the global
$AdS_5$ coordinate by 
\be
\tan\phi={\tan\eta\over\sin\mu}\ ,\qquad\qquad
\cos\theta=\cos\mu\cos\eta\ ,
\ee
where $0\le\phi\le\pi$ and $0\le\theta\le\pi/2$.
In terms of $\phi$ and $\theta$ the metric becomes
\be
ds^2={1\over\cos^2\theta}\left(-d\tau^2+d\theta^2+\sin^2\theta
\left(d\phi^2+\sin^2\phi d\Omega_2^2\right)\right)\ .
\ee

The boundary of $AdS_5$ is at $\theta=\pi/2$ which corresponds to 
$\mu=\pm\pi/2$ or $\eta=\pi/2$.
At $\mu=\pm\pi/2$, $\tan\phi=\pm\tan\eta$ respectively.
Since the range of $\eta$ is from $0$ to $\pi/2$,  $\tan\phi>0$ for
$\mu=+\pi/2$ and $\tan\phi<0$ for $\mu=-\pi/2$. 
This means that $0\le\phi<\pi/2$ for $\mu=+\pi/2$ and
$\pi/2<\phi\le\pi$ for $\mu=-\pi/2$.
Hence $\mu=+\pi/2$ corresponds to the upper hemi-sphere and
$\mu=-\pi/2$ to the lower hemi-sphere of $S^3$, and they are joined at
$\eta=\pi/2$. 

\begin{figure}[ht!]
\centering \epsfysize=7cm
\includegraphics[scale=0.3]{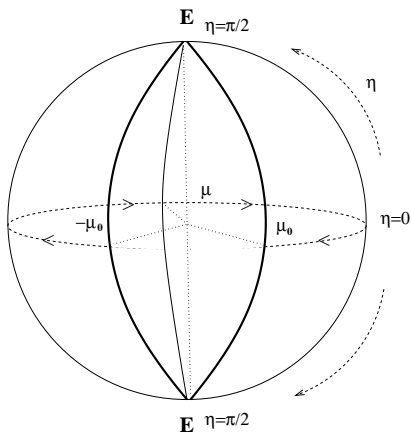}
\caption{The conformal mapping of the spatial section of Janus in the
  global coordinates. The two dimensional surface shown is
parametrized by $(\mu,\eta)$, and has the topology of disk. 
Each point in the upper half of the surface is identified with that in
the lower half due to the rotational symmetry associated with  
$S^2$ whose $(\mbox{radius})^2\propto \sin^2\eta$. 
The boundary is indicated by the thick line corresponding to
$S^3=S^3_+(\mu=\mu_0)\cup S^2(\mbox{E})\cup S^3_-(\mu=-\mu_0)$, 
where E denotes the equator of $S^3$.} 
\label{global}
\end{figure}

Qualitatively the same is true for the deformed case. The only
difference is that $\mu$ now ranges from $-\mu_0$ to $+\mu_0$ with
$\mu_0>\pi/2$. Thus $\mu=+\mu_0$ corresponds to the upper hemi-sphere 
$S^3_+$ and $\mu=-\mu_0$ to the lower hemi-sphere $S^3_-$ of $S^3$,
and they are joined at 
$\eta=\pi/2$. This is depicted in Figure \ref{global}.

Similarly for the latter the coordinate transformation
\be
x=y\sin\mu\ ,\qquad\qquad
z=y\cos\mu\ ,
\ee
brings the metric (\ref{poincareAdS4}) into
\be
ds^2={1\over z^2}\left(-dt^2+d\vec{x}_2^2+dx^2+dz^2\right)\ .
\ee

The boundary of $AdS_5$ is at $z=0$ which corresponds to
$\mu=\pm\pi/2$ or $y=0$.
At $\mu=\pm\pi/2$, $x=\pm y$ respectively.
Since $y\ge 0$, $x>0$ for $\mu=+\pi/2$ and $x<0$ for $\mu=-\pi/2$
except at $y=0$.
Hence $\mu=\pm\pi/2$ each corresponds to a half
$\mathbb{R}^{3,1}_{\pm}$ of the boundary $\mathbb{R}^{3,1}$, 
and they are joined at $y=0$.

\begin{figure}[ht!]
\centering \epsfysize=7cm
\includegraphics[scale=0.4]{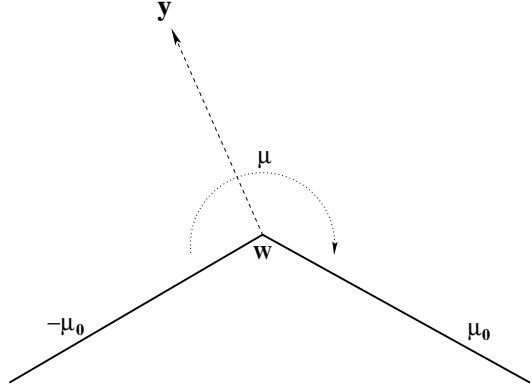}
\caption{The conformal mapping of the spatial section of Janus in the
  Poincare patch. The plane is parametrized by $(\mu,y)$.
Each point on the plane corresponds to $\mathbb{R}^2$.
The boundary is indicated by the wedge with thick line 
corresponding to $\mathbb{R}^3=\mathbb{R}^3_+(\mu=\mu_0)\cup
\mathbb{R}^2(\mbox{W})\cup\mathbb{R}^3_-(\mu=-\mu_0)$.}
\label{poincare}
\end{figure}

Again qualitatively the same is true for the deformed case. Only
difference from the undeformed case is the range of $\mu$. 
This is shown in Figure \ref{poincare}.

This peculiar structure of the Janus geometry reveals an interesting
feature from the viewpoint of the AdS/CFT duality \cite{Bak:2003jk}.
Since the constant value of the dilaton differs in each half of the
boundary at $\mu=\pm\mu_0$, the SYM coupling jumps when it moves
from one half of the space to another, yielding two different values
of the coupling constants dividing the spacetime into two.  
That is, it is as if the dual gauge theory had two faces -- hence the
name Janus, the god of gates, doors, doorways, beginnings, and endings
in Roman mythology who is often symbolized by two faces.
Two faces are joined at the interface. Although the 4-dimensional
conformal symmetry $SO(2,4)$ is partially broken, the conformal
symmetry $SO(2,3)$ on the interface is preserved. 
Hence the dual gauge theory is thought of as an interface CFT
\cite{Clark:2004sb}.

\section{The nested Janus}
Since the $AdS_4$-slice can be further sliced into $AdS_3$ and even
further into $AdS_2$, it may be possible to nest little Janus into the
interface of the parent Janus in a self-similar way.
In terms of the dual gauge theory, this would correspond to having an
interface CFT of higher codimensions.
In this section, we will show that this is indeed the case.
It turns out that the result is as simple as it could be.

The ansatz for the nested Janus solution takes the form
\bea
ds^2&=&f_1(\mu_1)\Biggl(d\mu_1^2+f_2(\mu_2)\Bigl(d\mu_2^2+
f_3(\mu_3)\left(d\mu_3^2+ds_{AdS_2}^2\right)\Bigr)\Biggr)
+ds_{S^5}^2\ ,\nn\\
\phi&=&\phi_1(\mu_1)+\phi_2(\mu_2)+\phi_3(\mu_3)\ ,
\label{nestedJ}\\
F_5&=&4f_1(\mu_1)^{5/2}f_2(\mu_2)^2f_3(\mu_3)^{3/2}
d\mu_1\wedge d\mu_2\wedge d\mu_3\wedge d\omega_{AdS_2}+4d\omega_{S^5}\
.\nn 
\eea
The isometry of the geometry is $SO(2,1)\times SO(6)$ in generic
cases. It could be enhanced to larger isometries in special
cases.

The equation of motion for the dilaton yields
\be
f_2^2f_3^{3/2}\left(f_1^{3/2}\phi_{1,1}\right)_{,1}
+f_1^{3/2}f_3^{3/2}\Bigl(f_2\phi_{2,2}\Bigr)_{,2}
+f_1^{3/2}f_2\Bigl(f_3^{1/2}\phi_{3,3}\Bigr)_{,3}=0\ ,
\label{dileqs}
\ee
where the symbol $h_{,i}$ for any function $h$ denotes the derivative
of $h$ with respect to $\mu_i$. It can be solved by
\be
\phi_{1,1}={c_1\over f_1^{3/2}}\ ,\qquad
\phi_{2,2}={c_2\over f_2}\ ,\qquad
\phi_{3,3}={c_3\over f_3^{1/2}}\ .
\label{nestdilaton}
\ee
The Einstein equations then yield
\bea
2f_{1,1}f_{1,1}-2f_1f_{1,1,1}&=&
-4f_1^3+{c_1^2\over 2f_1}\ ,\nn\\
-\frac{f_{1,1}f_{1,1}+2f_1f_{1,1,1}}{f_1^2}
+\frac{6f_{2,2}f_{2,2}-6f_2f_{2,2,2}}{f_2^3}&=&
-16f_1+{2c_2^2\over f_2^3}\ ,\nn\\
-\frac{f_{1,1}f_{1,1}+2f_1f_{1,1,1}}{f_1^2}
-{2f_{2,2,2}\over f_2^2}
+\frac{4f_{3,3}f_{3,3}-4f_3f_{3,3,3}}{f_2f_3^3}
&=&-16f_1+{2c_3^2\over f_2f_3^2}\ ,\nn\\
-{f_{1,1}f_{1,1}+2f_1f_{1,1,1}\over f_1^2}
-{2f_{2,2,2}\over f_2^2}
+{f_{3,3}f_{3,3}-2f_3f_{3,3,3}-4f_3^2\over f_2f_3^3}
&=&-16f_1\ .\nn
\eea
By inspection it turns out that these equations can be solved by
\bea
f_{1,1}f_{1,1}+2f_1f_{1,1,1}+12f_1^2-16f_1^3&=&0\ ,\nn\\
f_2f_{2,2,2}+4f_2^2-6f_2^3&=&0\ ,\nn\\
f_{3,3}f_{3,3}-2f_3f_{3,3,3}-4f_3^2+8f_3^3&=&0\ ,\nn\\
2f_{1,1}f_{1,1}-2f_1f_{1,1,1}+4f_1^3-{c_1^2\over 2f_1}&=&0\ ,\nn\\
3f_{2,2}-2f_2f_{2,2,2}+4f_2^2-6c_2^2&=&0\ ,\nn\\
3f_{3,3}f_{3,3}-2f_3f_{3,3,3}+4f_3^2-2c_3^2f_3&=&0\ .\nn
\eea
Note that three scale factors $f_{i=1,2,3}$ do not mix in the
equations. Similarly to the Janus solution reviewed in the previous
section, it is easy to show that these equations are equivalent to the
particle motion governed by the Hamiltonians
\bea
H_1(f_1,f_{1,1})&\equiv&\half f_{1,1}f_{1,1}+V_1(f_1)
=\half f_{1,1}f_{1,1}-\half\left(4f_1^3-4f_1^2
+{c_1^2\over 6f_1}\right)\ ,
\label{f1hamilton}\\
H_2(f_2,f_{2,2})&\equiv&\half f_{2,2}f_{2,2}+V_2(f_2)
=\half f_{2,2}f_{2,2}-\half\left(4f_2^3-4f_2^2+2c_2^2\right)
\ ,
\label{f2hamilton}\\
H_3(f_3,f_{3,3})&\equiv&\half f_{3,3}f_{3,3}+V_3(f_3)
=\half f_{3,3}f_{3,3}-\half\left(4f_3^3-4f_3^2+c_3^2f_3\right)
\ ,\label{f3hamilton}
\eea
with zero energies $H_i(f_i,f_{i,i})=0$ ($i=1,2,3$).

\begin{figure}[ht]
\begin{center}
$\begin{array}{c@{\hspace{1in}}c}
\multicolumn{1}{l}{\mbox{ }} &
        \multicolumn{1}{l}{\mbox{ }} \\ [-0.6cm]
\epsfxsize=2.2in
\epsffile{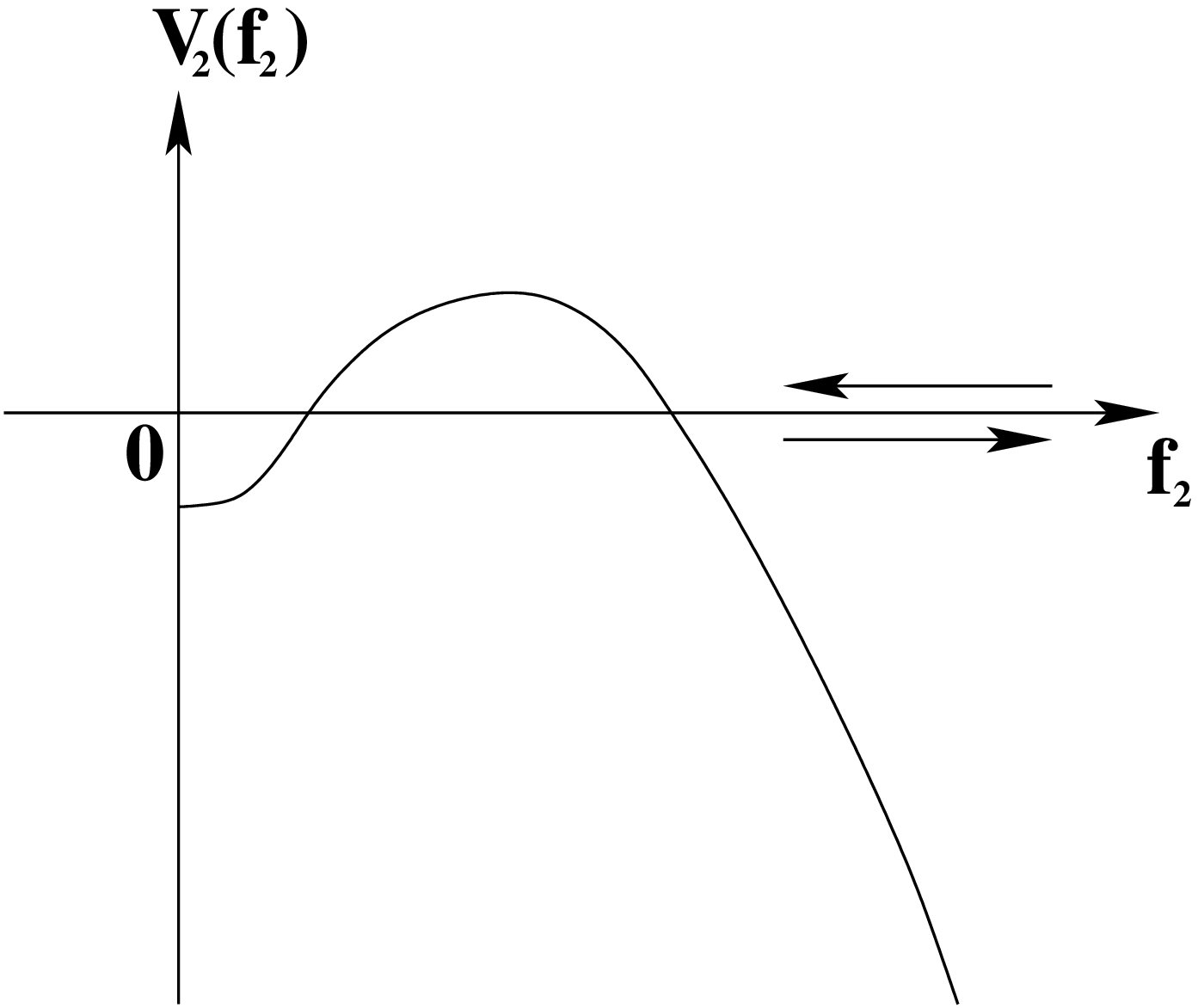} &
        \epsfxsize=2in
        \epsffile{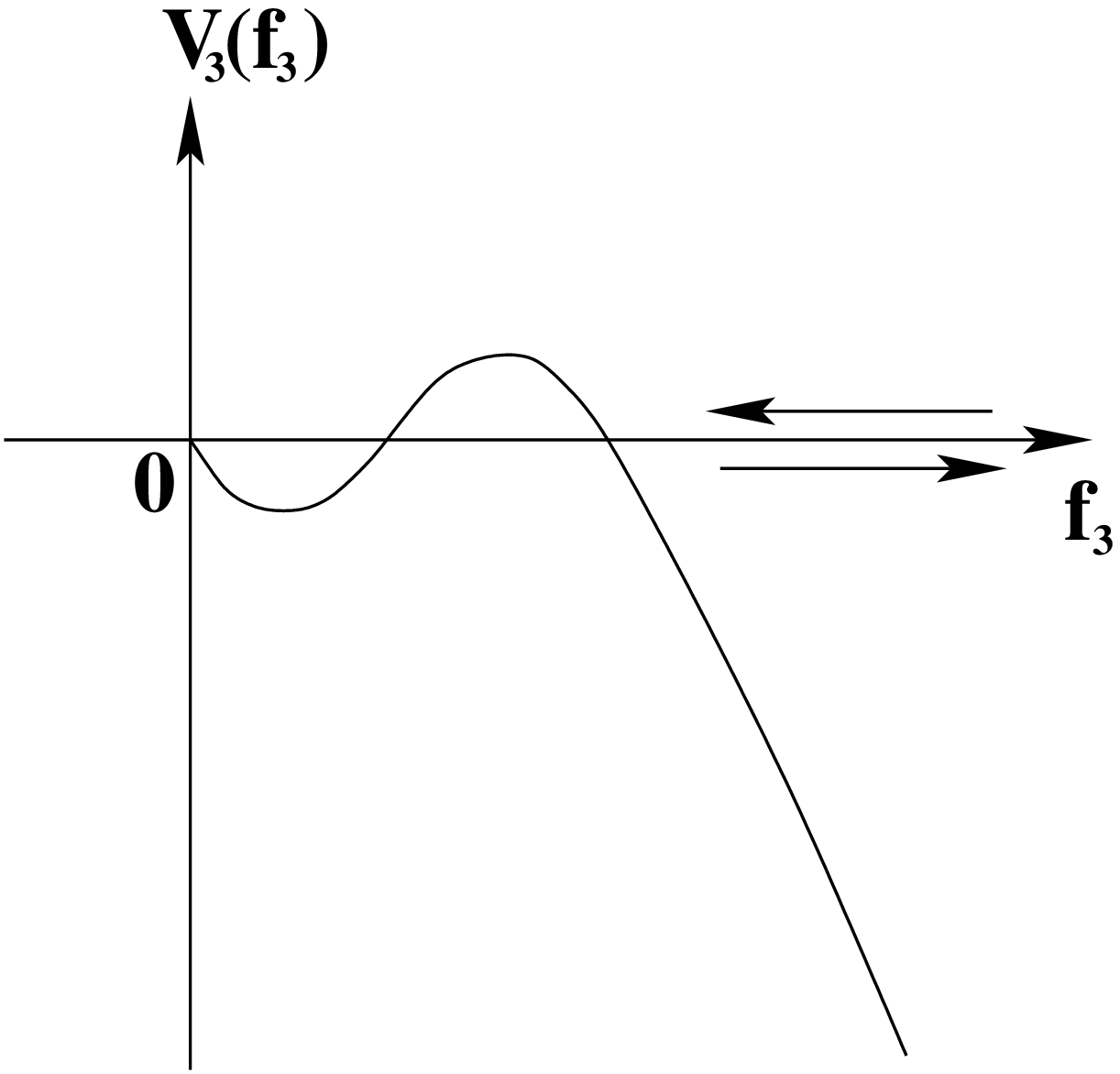} \\ [0.1cm]
\mbox{\bf (A)} & \mbox{\bf (B)}
\end{array}$
\end{center}
\caption{The motions of a particle governed (A) by the Hamiltonian
  (\ref{f2hamilton}) and (B) by (\ref{f3hamilton}) with zero energy,
  corresponding to the $AdS_3$ and $AdS_2$-sliced Janus respectively.}  
\label{potential2}
\end{figure}

The scale factor $f_1(\mu_1)$ obeys exactly the same equation as that
of the $AdS_4$-sliced Janus in the previous section,
corresponding to the particle motion depicted in Figure
\ref{potential}.  
More interestingly the scale factors $f_2(\mu_2)$ and $f_3(\mu_3)$ 
obey respectively the equation for the $AdS_2$ and $AdS_3$-sliced
Janus generalized in \cite{Freedman:2003ax}, corresponding to the
particle motions shown in Figure \ref{potential2}.

Hence the geometry represents Janus within Janus -- $AdS_2$-sliced
Janus $\subset$ $AdS_3$-sliced Janus $\subset$ $AdS_4$-sliced Janus.  
Little Janus inside the parent Janus follows the remarkably simple
``nesting'' rule in a self-smilar fashion, and the dilaton follows the
simple sum rule.  

As a further illustration, let us consider a simplified example -- the
$AdS_4$ slicing of the undeformed $AdS_5$ where the $AdS_4$-slice is
further sliced by $AdS_3$.
In the Poincare patch the metric takes the form
\be
ds^2={1\over\cos^2\mu_1}\left(d\mu_1^2+{1\over y_2^2\cos^2\mu_2}
\left(-dt^2+dx_3^2+dy_2^2+y_2^2d\mu_2^2\right)\right)\ ,
\label{AdS3poincare}
\ee
where $-\pi/2\le\mu_{1,2}\le\pi/2$ and $y_2\ge 0$.

We will transform it to the standard form of the Poincare patch for
$AdS_5$ in two steps.
 
In the first step, we introduce the new coordinates by
\be
x_2=y_2\sin\mu_2\ ,\qquad\qquad
y_1=y_2\cos\mu_2\ .
\ee
This brings (\ref{AdS3poincare}) into the form
\be
ds^2={1\over\cos^2\mu_1}\left(d\mu_1^2+{1\over y_1^2}
\left(-dt^2+dx_3^2+dx_2^2+dy_1^2\right)\right)\ ,
\label{poincareint}
\ee
where $y_1\ge 0$.

Similarly in the next step, by the coordinate transformation
\be
x_1=y_1\sin\mu_1\ ,\qquad\qquad
z=y_1\cos\mu_1\ ,
\ee
the metric (\ref{poincareint}) yields
\be
ds^2={1\over z^2}\left(-dt^2+dx_3^2+dx_2^2+dx_1^2+dz^2\right)\ , 
\label{AdS4AdS3poincare}
\ee
where $z\ge 0$.

The coordinate transformation we made from $(t,x_3,\mu_1,\mu_2,y_2)$
to $(t,x_3,x_2,x_1,z)$ is
\be
z=y_2\cos\mu_1\cos\mu_2\ ,\qquad
x_1=y_2\sin\mu_1\cos\mu_2\ ,\qquad
x_2=y_2\sin\mu_2\ .
\label{transformation}
\ee

Now the boundary of $AdS_5$ is at $z=0$ which corresponds to 
$\mu_1=\pm\pi/2$, $\mu_2=\pm\pi/2$, or $y_2=0$.
At $\mu_2=\pm\pi/2$, $x_1=0$ and $x_2=\pm y_2$ respectively. 
Since $y_2\ge 0$, $x_2>0$ for $\mu_2=+\pi/2$ and $x_2<0$ for
$\mu_2=-\pi/2$ except at $y_2=0$.
Hence $\mu_2=\pm\pi/2$ each corresponds to a half
$\mathbb{R}^{2,1}_{\pm}$ of 
the codimension 1 subspace $\mathbb{R}^{2,1}$ (defined by $x_1=0$) of
the boundary $\mathbb{R}^{3,1}$, and they are joined at $y_2=0$ which
is the codimension 2 subspace $\mathbb{R}^{1,1}$ (defined by
$x_1=x_2=0$).  
Similarly at $\mu_1=\pm\pi/2$, $x_1=\pm y_2\cos\mu_2$ respectively.
Since $y_2\ge 0$ and $-\pi/2\le\mu_2\le\pi/2$, $x_1>0$ for $\mu_1=\pi/2$
and $x_1<0$ for $\mu_1=-\pi/2$ except at $y_2=0$ or $\mu_2=\pm\pi/2$.
Hence $\mu_1=\pm\pi/2$ each corresponds to a half
$\mathbb{R}^{3,1}_{\pm}$ of the boundary $\mathbb{R}^{3,1}$ 
joined at the subspace $\mathbb{R}^{2,1}$ defined by 
$y_2(\mu_2+\pi/2)(\mu_2-\pi/2)=0$.

As in the simpler Janus in the previous section, qualitatively the
same is true for the deformed case. Only difference is the range of
$\mu_{1,2}$, instead of $-\pi/2\le\mu_{1,2}\le\pi/2$, we have 
$-\mu_{1,0}\le\mu_1\le\mu_{1,0}$ and
$-\mu_{2,0}\le\mu_2\le\mu_{2,0}$ where $\mu_{i,0}\ge\pi/2$.
 
\begin{figure}[ht!]
\centering \epsfysize=6cm
\includegraphics[scale=0.45]{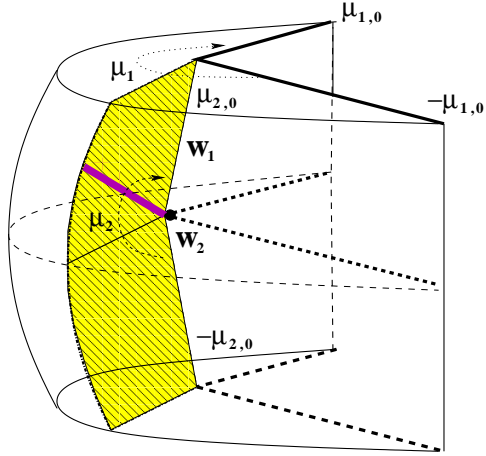}
\caption{The $AdS_4$-sliced Janus and $AdS_3$-sliced Janus within
  it in the Poincare patch. The patterned yellow area is a (constant
  $\mu_1$) slice of the (deformed) $AdS_4$. The thick magenta line in
  it is a (constant $\mu_2$) slice of the (deformed) $AdS_3$.
The wedge $W_1$ depicted as a vertical line is 
$\mathbb{R}^{2,1}=\mathbb{R}^{2,1}_+
\cup\mathbb{R}^{1,1}\cup\mathbb{R}^{2,1}_-$ 
  which is the interface of two faces of the $AdS_4$-sliced Janus. 
The wedge $W_2$ depicted as a point is 
$\mathbb{R}^{1,1}=\mathbb{R}^{1,1}_+
\cup\mathbb{R}^{1}\cup\mathbb{R}^{1,1}_-$
  corresponding to the interface of two faces of the $AdS_3$-sliced
  Janus nested in the $AdS_4$-sliced Janus. 
Upon the inclusion of the $AdS_2$-sliced Janus,
  the timelike $\mathbb{R}^{1}\subset\mathbb{R}^{1,1}$ would
  correspond to its interface.  }  
\label{slice}
\end{figure}

It should also be clear that the $AdS_3$-sliced Janus nests in the
codimension 1 subspace $\mathbb{R}^{2,1}$ which is nothing but the
interface of the $AdS_4$-sliced Janus -- hence the name nested Janus.    
This structure is depicted in Figure \ref{slice}.
It is straightforward to extend this argument further to the $AdS_2$
slicing.\footnote{In the case of the global coordinates, the boundary
  of $AdS_2$ consists of two disconnected components
  corresponding to $\eta=\pm\pi/2$ in (\ref{globalAdS4}) if applied
  to the $AdS_2$ space {\it per se}. 
However, since $\eta$ ranges from $0$
  to $\pi/2$ for the $AdS_2$ slice in $AdS_5$, one of the components
  is singled out for the $AdS_2$-sliced Janus.
We would like to thank M. Gutperle for asking us to clarify this
point.}    
So we will not repeat it here.

A few remarks are in order: 
First, the equations $H_i(f_i,f_{i,i})=0$ ($i=1,2,3$) as well as the
dilaton equations (\ref{dileqs}) can be easily
integrated as in the case of the simple Janus in the previous
section. 
As mentioned before, the analytic form of the function $f_1(\mu_1)$
was recently found in \cite{D'Hoker:2006uu} in terms of the
Weierstrass $\wp$-function. 
That of $f_3(\mu_3)$ is simpler and the expicit form is given in
\cite{Freedman:2003ax} in terms of the Jacobi elliptic function.  
On the other hand, that of $f_2(\mu_2)$ is yet to be found.

Regardless of the explicit form of the scale factors, the qualitative
feature of each $AdS_{4,3,2}$-sliced Janus is essentially all the same 
-- their boundaries consist of two parts joined at the interface, and 
the constant value of the dilaton differs in each half of the boundary
at $\mu_i=\pm\mu_{i,0}$  
where $\phi_i(\pm\mu_{i,0})$ takes the value
$\phi_{i,0}\pm\Delta\phi_{i,0}$ respectively.

Second, as in the case of the $AdS_4$-sliced Janus, there are
upper bounds on the constants $c_{2,3}$ for the $AdS_{2,3}$-sliced
Janus, above which the non-singular solution ceases to exist. As can
be easily seen from the form of the potential in (\ref{f2hamilton})
and (\ref{f3hamilton}), as we increase the value of $c_{2,3}$, the
potentials go down, and at some values of $c_{2,3}$ the top of the
potentials is leveled with $V_{2,3}=0$.  
It is easy to find that it occurs at $c_2=(2/3)^{3/2}$ and 
$c_3=1$.
Hence we are only interested in the value of $c_{1,2,3}$ in the ranges
\bea
0\le &c_1&\le 9/4\sqrt{2}\ ,\nn\\
0\le &c_2&\le (2/3)^{3/2}\ ,\nn\\
0\le &c_3&\le 1\ .\nn
\eea

Third, it is clear that, for example, the $AdS_2$-sliced Janus alone
can exist in $AdS_5$ without nesting. This is the codimension 3
Janus. 
Similarly the $AdS_3$-sliced Janus alone corresponds to
the codimension 2 Janus.
Also any combination of two is obviously possible.

Finally the boundary of the nested Janus is at $f_i(\mu_i)=+\infty$
($i=1,2,3$) or equivalently at $\mu_i=\pm\mu_{i,0}$ 
where the geometry asymptotes to $AdS_5$. 
The asymptotics of the nested Janus solution can be easily found as 
\be
f_i(\mu_i)\sim {1\over (\mu_i\mp\mu_{i,0})^2}\ ,
\ee
\bea
\phi(\mu_1,\mu_2,\mu_3)&\sim&
\phi_0\pm\Delta\phi_{1,0}\mp{c_1\over 4}(\mu_1\mp\mu_{1,0})^4 \nn\\
&&
\pm\Delta\phi_{2,0}+{c_2\over 3}(\mu_2\mp\mu_{2,0})^3
\pm\Delta\phi_{3,0}\mp{c_3\over 2}(\mu_3\mp\mu_{3,0})^2\ ,
\label{asympdil}
\eea
where we have defined $\phi_0=\sum_{i=1,3}\phi_{i,0}$.

In the dual gauge theory, as proposed in \cite{Bak:2003jk} and refined
in \cite{Clark:2004sb}, the SYM coupling discontinuously jumps
when it crosses the interface of two halves of the plane.
In the nested Janus case, the interface accommodates
lower dimensional interfaces within itself,   
so there is a sequence of jumps of the gauge coupling.
We propose that the deformation to add to ${\cal N}=4$ SYM is 
\be
\Delta S=-\int d^4x\Bigl(\Delta\gamma_1\varepsilon(x_1)
+\Delta\gamma_2\delta^{\epsilon}_{x_1,0}\varepsilon(x_2)
+\Delta\gamma_3\delta^{\epsilon}_{(x_1,0),(x_2,0)}\varepsilon(x_3)
\Bigr){\cal L}_{SYM}\ ,
\ee
where ${\cal L}_{SYM}$ is the Lagrangian density of ${\cal N}=4$ SYM
with the difference by a total derivative term \cite{Clark:2004sb}, 
and the step function $\varepsilon(x)=2\theta(x)-1$.
We have also defined the functions $\delta^{\epsilon}_{x_1,0}\equiv{\epsilon^2\over\epsilon^2+x_1^2}$ 
and $\delta^{\epsilon}_{(x_1,0),(x_2,0)}\equiv{\epsilon^4\over(\epsilon^2+x_1^2+x_2^2)^2}$, where $\epsilon$ should be taken to zero after all the calculations.
The undeformed action is given by ${1\over g_{YM}^2}{\cal L}_{SYM}$
with 
$1/g_{YM}^2=\prod_{i=1}^3(1/g_i^2)\equiv
\prod_{i=1}^3\left(\half(1/g_{i+}^2+1/g_{i-}^2)\right)$ 
where $g_{i\pm}^2=2\pi e^{\phi_{i,0}\pm\Delta\phi_{i,0}}$.
The deformation parameters $\Delta\gamma_i$ are related to $g_{i\pm}$s
by  
\be
\Delta\gamma_i={1\over g_{YM}^2}\,\,{g_{i+}^2-g_{i-}^2\over
  g_{i+}^2+g_{i-}^2}\ .
\ee
This deformation will induce the vev for the dimension 4 operator
${\cal L}_{SYM}$,  
as expected from the subleading contribution in the asymptotic
expansion of the dilaton.

To see it, 
let us take a closer look at the asymptotic expansion (\ref{asympdil})
of the dilaton. 
Since we wish to consider the dual gauge theory on
$\mathbb{R}^{3,1}$, we need to know the relation between two
coordinate systems, the $AdS$-slicings $(t,\mu_1,\mu_2,\mu_3,y_3)$ 
and the $AdS_5$ Poincare patch $(t,x_1,x_2,x_3,z)$, where $(t,y_3)$ is
the coordinates of the $AdS_2$ Poincare patch.
Near $\mu_i=\pm\mu_{i,0}$ the nested Janus is approximately $AdS_5$, 
so similarly to the transformation (\ref{transformation}) discussed
above, we can find that near the boundary 
\bea
&&z\sim \pm y_3\sin(\mu_1\pm\mu_{1,0})\sin(\mu_2\pm\mu_{2,0})
\sin(\mu_3\pm\mu_{3,0})\ ,\nn\\
&&x_1\sim \mp y_3\cos(\mu_1\pm\mu_{1,0})\sin(\mu_2\pm\mu_{2,0})
\sin(\mu_3\pm\mu_{3,0})\ ,\nn\\
&&x_2\sim - y_3\cos(\mu_2\pm\mu_{2,0})\sin(\mu_3\pm\mu_{3,0})
\ ,\nn\\
&&x_3\sim \mp y_3\cos(\mu_3\pm\mu_{3,0})\ , \nn
\eea
where the order of signs are correlated.
We have fixed the signs as follows: First recall that $\mu_i$s are in
the ranges $-\mu_{i,0}\le\mu_i\le\mu_{i,0}$ and we are considering
$\mu_i\sim\pm\mu_{i,0}$. So we have $\sin(\mu_i+\mu_{i,0})>0$, 
$\sin(\mu_i-\mu_{i,0})<0$, and $\cos(\mu_i\pm\mu_{i,0})>0$. Now since
$z\ge 0$, that fixes the sign in the first line.
The rest is fixed based on the fact that 
$x_i>0$ at $\mu_i=\mu_{i,0}$ and $x_i<0$ at $\mu_i=-\mu_{i,0}$.
 
We can then deduce that 
\bea
\tan(\mu_1\pm\mu_{1,0})&\sim& -{z\over x_1}\ ,\nn\\
\tan(\mu_2\pm\mu_{2,0})&\sim&\pm
\sqrt{{z^2+x_1^2\over x_2^2}}\ ,\nn\\
\tan(\mu_3\pm\mu_{3,0})&\sim&
\pm\sqrt{{z^2+x_1^2+x_2^2\over x_3^2}}\ .\nn
\eea
Hence we obtain
\bea
(\mu_1\pm\mu_{1,0})^4&\sim&
\left({z\over x_1}\right)^4\ ,\\
(\mu_2\pm\mu_{2,0})^3&\sim&\pm
\left({z^2+x_1^2\over x_2^2}\right)^{3/2}
=\pm\left({z\over z^2+x_1^2}\right)
{z^2+x_1^2\over z}\left({z^2+x_1^2\over x_2^2}\right)^{3/2}\nn\\
&\sim&\pm\pi\delta(x_1)
{z^2+x_1^2\over z}\left({z^2+x_1^2\over x_2^2}\right)^{3/2}
\sim -\pi\delta(x_1){z^4\over x_2^3}\ ,\\
(\mu_3\pm\mu_{3,0})^2&\sim&
{z^2+x_1^2+x_2^2\over x_3^2}
=\left({z^2\over (z^2+x_1^2+x_2^2)^2}\right)
{(z^2+x_1^2+x_2^2)^2\over z^2}{z^2+x_1^2+x_2^2\over x_3^2}\nn\\
&\sim&\pi\delta(x_1)\delta(x_2)
{(z^2+x_1^2+x_2^2)^2\over z^2}{z^2+x_1^2+x_2^2\over x_3^2}
\sim\pi\delta(x_1)\delta(x_2){z^4\over x_3^2}\ .
\eea
Strictly speaking, the order of limits we prescribed here 
needs to be justified more rigorously. 
However, the result so obtained appears to make
sense from the viewpoint of AdS/CFT.

Therefore according to the AdS/CFT dictionary \cite{Balasubramanian:1998sn}, 
what we would expect is
\be
\langle{\cal L}_{SYM}\rangle
=\varepsilon(x_1){N^2\over 2\pi^2}{c_1\over 4x_1^4}
+\delta(x_1)\varepsilon(x_2){N^2\over 2\pi}{c_2\over 3|x_2|^3}
+\delta(x_1)\delta(x_2)\varepsilon(x_3)
{N^2\over 2\pi}{c_3\over 2x_3^2}\ ,
\ee
generalizing the result of \cite{Clark:2004sb, Bak:2003jk} to the case
of nonvanishing $c_2$ and $c_3$.

\section{Discussions}

What makes Janus potentially interesting is its remarkable
simplicity. 
Here we have seen that this trait persists in the generalization we
discussed.  
The simple ``nesting'' rule found here is reminiscent of the
intersection rule for the supersymmetric brane configurations. 
So this might be related to the fact that the Janus solution has the
fake supersymmetry \cite{Freedman:2003ax, Skenderis:2006jq} even
though it is not a supersymmetric geometry in the standard
sense.\footnote{The ``fake'' supersymmetry constraint on the $AdS$
  domain wall was first derived in \cite{Behrndt:2002ee}, and its
  relation to the real supergravity was estabilished in   
\cite{Celi:2004st}.} 

The stability of the nested Janus solution needs to be examined.  
However, since the $AdS_d$-sliced Janus for any dimension $d$ was
shown to be stable against a large class of perturbations  
\cite{Freedman:2003ax}, we believe that this is also the case for the
nested Janus which is made up of a sequence of the $AdS_d$-sliced
Janus each one of which is stable.

It is worthwhile to study the dual interface CFT and compare the
results with the supergravity predictions, generalizing the firm
analysis carried out in \cite{Clark:2004sb}. 
The holographic renormalization group method developed in
\cite{Papadimitriou:2004rz} enables more efficient computations of the
correlation functions. In this paper we computed only the vev of a
particular dimension 4 operator, but it is worth calculating the
correlation functions by using their method to check with the
interface CFT expectation.
In particular, since the $AdS_2$-sliced Janus is simpler relative to
the higher dimensional counterparts, it might be useful to study this
case more in details. 

Finally it must be possible to find the supersymmetric version of
the nested Janus.
In the dual interface CFT side, there must exist the
interface interactions which restore the supersymmetries
\cite{Clark:2004sb, D'Hoker:2006uv}. 
Perhaps the thorough classification, as was done in the case of the
supersymmetric Janus in \cite{D'Hoker:2006uv}, of such
interactions in this more general case would in turn suggest whether
and how the supersymmetrization can be made in type IIB
supergravity. 

\section*{Acknowledgements}

The author would like to thank Dongsu Bak and Michael Gutperle for
comments.



\begin{thebibliography}{40}

\bibitem{Bak:2003jk}
D.~Bak, M.~Gutperle and S.~Hirano,
``A dilatonic deformation of AdS(5) and its field theory dual,''
JHEP {\bf 0305}, 072 (2003)
[arXiv:hep-th/0304129].


\bibitem{Maldacena:1997re}
  J.~M.~Maldacena,
  ``The large N limit of superconformal field theories and supergravity,''
  Adv.\ Theor.\ Math.\ Phys.\  {\bf 2}, 231 (1998)
  [Int.\ J.\ Theor.\ Phys.\  {\bf 38}, 1113 (1999)]
  [arXiv:hep-th/9711200].


\bibitem{Gubser:1998bc}
  S.~S.~Gubser, I.~R.~Klebanov and A.~M.~Polyakov,
  ``Gauge theory correlators from non-critical string theory,''
  Phys.\ Lett.\ B {\bf 428}, 105 (1998)
  [arXiv:hep-th/9802109].

\bibitem{Witten:1998qj}
  E.~Witten,
  ``Anti-de Sitter space and holography,''
  Adv.\ Theor.\ Math.\ Phys.\  {\bf 2}, 253 (1998)
  [arXiv:hep-th/9802150].

\bibitem{Aharony:1999ti}
  O.~Aharony, S.~S.~Gubser, J.~M.~Maldacena, H.~Ooguri and Y.~Oz,
  ``Large N field theories, string theory and gravity,''
  Phys.\ Rept.\  {\bf 323}, 183 (2000)
  [arXiv:hep-th/9905111].

\bibitem{Kehagias:1999tr}
  A.~Kehagias and K.~Sfetsos,
  ``On running couplings in gauge theories from type-IIB supergravity,''
  Phys.\ Lett.\ B {\bf 454}, 270 (1999)
  [arXiv:hep-th/9902125].

\bibitem{Gubser:1999pk}
  S.~S.~Gubser,
  ``Dilaton-driven confinement,''
  arXiv:hep-th/9902155.

\bibitem{Nojiri:1999sb}
  S.~Nojiri and S.~D.~Odintsov,
  ``Curvature dependence of running gauge coupling and confinement in
  AdS/CFT correspondence,''
  Phys.\ Rev.\ D {\bf 61}, 044014 (2000)
  [arXiv:hep-th/9905200].

\bibitem{Bak:2004yf}
  D.~Bak, M.~Gutperle, S.~Hirano and N.~Ohta,
  ``Dilatonic repulsons and confinement via the AdS/CFT correspondence,''
  Phys.\ Rev.\ D {\bf 70}, 086004 (2004)
  [arXiv:hep-th/0403249]; 
  D.~Bak and H.~U.~Yee,
  ``Separation of spontaneous chiral symmetry breaking and confinement via
  AdS/CFT correspondence,''
  Phys.\ Rev.\ D {\bf 71}, 046003 (2005)
  [arXiv:hep-th/0412170].


\bibitem{Freedman:2003ax}
  D.~Z.~Freedman, C.~Nunez, M.~Schnabl and K.~Skenderis,
  ``Fake supergravity and domain wall stability,''
  Phys.\ Rev.\ D {\bf 69}, 104027 (2004)
  [arXiv:hep-th/0312055].

\bibitem{Clark:2004sb}
  A.~B.~Clark, D.~Z.~Freedman, A.~Karch and M.~Schnabl,
  ``The dual of Janus (($<:$) $<$--$>$ ($:>$)) an interface CFT,''
  Phys.\ Rev.\ D {\bf 71}, 066003 (2005)
  [arXiv:hep-th/0407073].


\bibitem{D'Hoker:2006uv}
  E.~D'Hoker, J.~Estes and M.~Gutperle,
  ``Interface Yang-Mills, Supersymmetry, and Janus,''
  arXiv:hep-th/0603013.


\bibitem{Karch:2000ct}
  A.~Karch and L.~Randall,
  ``Locally localized gravity,''
  JHEP {\bf 0105}, 008 (2001)
  [arXiv:hep-th/0011156].


\bibitem{Behrndt:2002ee}
  K.~Behrndt and M.~Cvetic,
  ``Bent BPS domain walls of D = 5 N = 2 gauged supergravity coupled
  to hypermultiplets,''
  Phys.\ Rev.\ D {\bf 65}, 126007 (2002)
  [arXiv:hep-th/0201272].


\bibitem{Sonner:2005sj}
  J.~Sonner and P.~K.~Townsend,
  ``Dilaton domain walls and dynamical systems,''
  Class.\ Quant.\ Grav.\  {\bf 23}, 441 (2006)
  [arXiv:hep-th/0510115].


\bibitem{Karch:2001cw}
  A.~Karch and L.~Randall,
  ``Localized gravity in string theory,''
  Phys.\ Rev.\ Lett.\  {\bf 87}, 061601 (2001)
  [arXiv:hep-th/0105108].

\bibitem{Clark:2005te}
  A.~Clark and A.~Karch,
  ``Super Janus,''
  JHEP {\bf 0510}, 094 (2005)
  [arXiv:hep-th/0506265].

\bibitem{D'Hoker:2006uu}
  E.~D'Hoker, J.~Estes and M.~Gutperle,
  ``Ten-dimensional supersymmetric Janus solutions,''
  arXiv:hep-th/0603012.

\bibitem{Bak:2006nh}
  D.~Bak,
  ``Dual of Big-bang and Big-crunch,''
  arXiv:hep-th/0603080.

\bibitem{Balasubramanian:1998sn}
  V.~Balasubramanian, P.~Kraus and A.~E.~Lawrence,
  ``Bulk vs. boundary dynamics in anti-de Sitter spacetime,''
  Phys.\ Rev.\ D {\bf 59}, 046003 (1999)
  [arXiv:hep-th/9805171].

\bibitem{Skenderis:2006jq}
  K.~Skenderis and P.~K.~Townsend,
  ``Hidden supersymmetry of domain walls and cosmologies,''
  arXiv:hep-th/0602260.

\bibitem{Celi:2004st}
  A.~Celi, A.~Ceresole, G.~Dall'Agata, A.~Van Proeyen and M.~Zagermann,
  ``On the fakeness of fake supergravity,''
  Phys.\ Rev.\ D {\bf 71}, 045009 (2005)
  [arXiv:hep-th/0410126].

\bibitem{Papadimitriou:2004rz}
  I.~Papadimitriou and K.~Skenderis,
  ``Correlation functions in holographic RG flows,''
  JHEP {\bf 0410}, 075 (2004)
  [arXiv:hep-th/0407071].

\end{thebibliography}
\end{document}